\documentclass[aip,reprint,apl]{revtex4-1}
\usepackage{graphicx}
\usepackage{dcolumn}
\usepackage{bm}
\usepackage{amsmath}
\usepackage{siunitx}

\begin{document}

\title{Frequency dependence of trapped flux sensitivity in SRF cavities} 

\author{M.~Checchin}
\email[]{checchin@fnal.gov}
\affiliation{Fermi National Accelerator Laboratory, Batavia IL 60510, USA}
\author{M.~Martinello}
\email[]{mmartine@fnal.gov}
\affiliation{Fermi National Accelerator Laboratory, Batavia IL 60510, USA}
\author{A.~Grassellino}
\affiliation{Fermi National Accelerator Laboratory, Batavia IL 60510, USA}
\affiliation{Department of Physics, Northwestern University, Evanston IL 60208, USA}
\author{S.~Aderhold}
\affiliation{Fermi National Accelerator Laboratory, Batavia IL 60510, USA}
\author{S.~K.~Chandrasekaran}
\affiliation{Fermi National Accelerator Laboratory, Batavia IL 60510, USA}
\author{O.~Melnychuk}
\affiliation{Fermi National Accelerator Laboratory, Batavia IL 60510, USA}
\author{S.~Posen}
\affiliation{Fermi National Accelerator Laboratory, Batavia IL 60510, USA}
\author{A.~Romanenko}
\affiliation{Fermi National Accelerator Laboratory, Batavia IL 60510, USA}
\affiliation{Department of Physics, Northwestern University, Evanston IL 60208, USA}
\author{D.~A.~Sergatskov}
\affiliation{Fermi National Accelerator Laboratory, Batavia IL 60510, USA}

\date{\today}

\begin{abstract}
In this letter, we present the frequency dependence of the vortex surface resistance of bulk niobium accelerating cavities as a function of different state-of-the-art surface treatments. Higher flux surface resistance per amount of trapped magnetic field\textemdash sensitivity\textemdash is observed for higher frequencies, in agreement with our theoretical model. Higher sensitivity is observed for N-doped cavities, which possess an intermediate value of electron mean-free-path, compared to $\SI{120}{\degreeCelsius}$ and EP/BCP cavities. Experimental results from our study showed that the sensitivity has a non-monotonic trend as a function of the mean-free-path, including at frequencies other than $1.3$~GHz, and that the vortex response to the rf field can be tuned from the pinning regime to flux-flow regime by manipulating the frequency and/or the mean-free-path of the resonator, as reported in our previous studies. The frequency dependence of the trapped flux sensitivity to the amplitude of the accelerating gradient is also highlighted.
\end{abstract}


\maketitle 

Superconducting radio-frequency (SRF) accelerating cavities are electromagnetic resonant structures employed in modern machines to accelerate charged particles to relativistic velocities. Machines operating in continuous wave (CW) have particularly an advantage in adopting the SRF technology, since the power dissipated during operation is lower due to the high Q-factors that SRF resonators can attain.

One of the main challenges in maintaining high Q-factors during the operation of a CW machine is the mitigation of the dissipations introduced by remnant magnetic field in the cryomodule. As separately studied by Martinello \textit{et al.}\cite{Martinello_APL_2016} and Gonnella \textit{et al.}\cite{Gonnella_JAP_2016}, the surface resistance of SRF cavities can be particularly sensitive to the magnetic field trapped during the cooldown of the resonator. It was indeed found that the sensitivity ($S$)\textemdash extra surface resistance introduced per amount of magnetic field trapped\textemdash at $1.3$~GHz is a non-monotonic function of the mean-free-path ($l$), and can reach a maximum value of about $1.5$~n$\Omega$/mG at approximately $l=70$~nm\cite{Martinello_APL_2016}. Such a peculiar mean-free-path dependence of the sensitivity is interpreted as the interplay of pinning and flux-flow dominated responses of the vortex dynamics to the rf field for small and large values of $l$, respectively\cite{Checchin_SUST_2017}.

The frequency dependence of the sensitivity is of crucial importance to unveil the vortex behavior under rf field, and it is also of great interest from the practical point of view. The cryomodule design and remnant magnetic field specifications for frequencies other than $1.3$~GHz are mostly set in the absence of experimental data on sensitivity, as of now. Thus, by means of the experimental findings presented in this letter we can then formulate guidelines for developing future SRF cryomodules that will adopt bulk niobium cavities operating at frequencies other than $1.3$~GHz.

In this letter, we present findings of our study on sensitivity to trapped flux as a function of cavity frequency by analyzing the trapped flux surface resistance of cavities operating at different frequencies and prepared with different surface treatments: electropolished (EP'd)\cite{Padamsee_Book2,Padamsee_AnnRevNucl_2014}, buffer chemical polished (BCP'd)\cite{Padamsee_Book2,Padamsee_AnnRevNucl_2014}, $\SI{120}{\degreeCelsius}$ baked\cite{Padamsee_Book2,Padamsee_AnnRevNucl_2014}, and N-doped\cite{Grassellino_SUST_2013}. Experimental data for elliptical cavities operating at $650$~MHz, $1.3$~GHz, $2.6$~GHz, and $3.9$~GHz are reported and compared with a theoretical model\cite{Checchin_SUST_2017} of the frequency dependence of $S$. The cryogenic rf test of cavities was conducted at the vertical test facility of the Fermi National Accelerator Laboratory, while the experimental setup and the sensitivity calculation procedure are reported in our previous work\cite{Martinello_APL_2016}.
\begin{figure*}[t]
\centering
\includegraphics[scale=1]{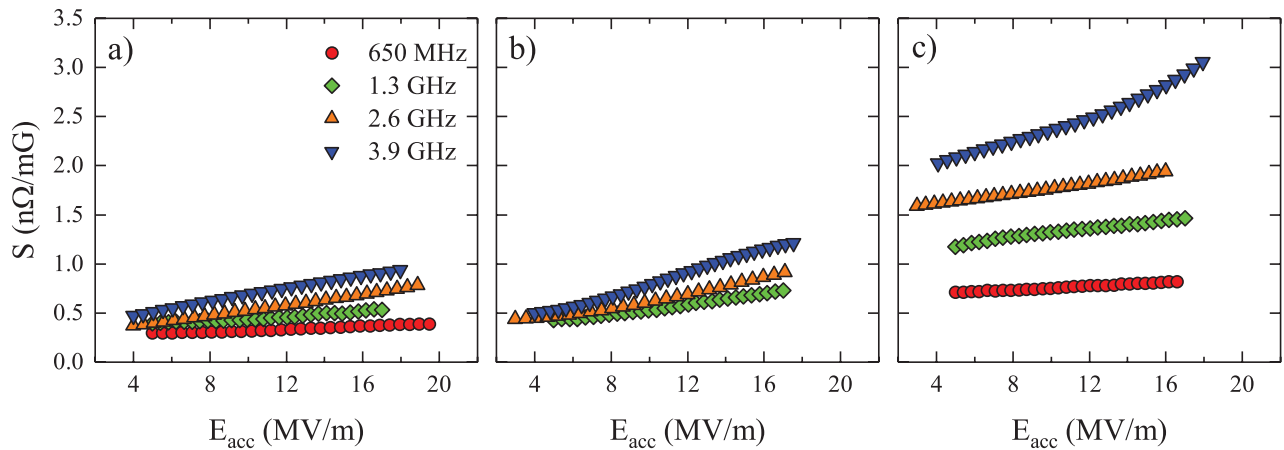}
\caption{\label{fig:Data}Sensitivity as a function of the accelerating gradient measured for the cavities studied. Graphs a), b), and c) show the data acquired for cavities treated with $\SI{120}{\degreeCelsius}$, EP/BCP, and 2/6 N-doping, respectively.}
\end{figure*}

Table~\ref{tab:mfp} presents the mean-free-path values and corresponding errors for the cavities studied. Unless otherwise specified, the mean-free-path values were extrapolated by fitting the frequency variation as a function of temperature (``$f$ vs $T$'' data) during the cavity warm-up by means of a C++ routine based on the Halbritter's code\cite{Halbritter_SRIMP_1970}, as described in Ref.~[\onlinecite{Martinello_APL_2016}]. 

The cavities treated with N-doping were baked with the recipe denominated as 2/6, selected for the energy upgrade of the Linear Coherent Light Source (LCLS-II) at SLAC\cite{Gonnella_JAP_2015,Liepe_LCLSII_SRF_2015,XFEL2}. The cavity was initially treated at $\SI{800}{\degreeCelsius}$ for 3 h to degas hydrogen. Nitrogen was then inlet with a partial pressure of $25$~mTorr,  the supply was closed after 2 min, and the cavity was annealed at the same temperature for additional 6 min.

In the graphs shown in Fig.~\ref{fig:Data}, measured data for the cavities listed in Table~\ref{tab:mfp} are reported. Figure~\ref{fig:Data}a, b, and c show respectively the sensitivity measured at different frequencies for $\SI{120}{\degreeCelsius}$ baked, EP'd/BCP'd, and N-doped cavities. EP and BCP surface finishing are discussed together because the respective values of the mean-free-path extracted from ``$f$ vs $T$'' measurements are comparable\textemdash as also confirmed by means of low-energy $\mu$SR data\cite{Romanenko_AppPhysLett_2014}\textemdash as well as their vortex response to rf field\cite{Martinello_APL_2016,Checchin_SUST_2017}. In all graphs, it is clear that the higher the frequency, the larger the sensitivity to trapped field, independent of the surface treatment. N-doped cavities showed $S$ twice higher (or more) than both EP'd/BCP'd and $\SI{120}{\degreeCelsius}$ baked cavities independent of the cavity frequency, which is in agreement with the expected non-monotonic behavior of the sensitivity observed at $1.3$~GHz\cite{Martinello_APL_2016}.
\begin{table*}[t]
\caption{\label{tab:mfp}Summary of the extrapolated mean-free-path values of the cavities studied. Columns and rows represent different frequencies and different surface treatments, respectively.}
\begin{ruledtabular}
\begin{tabular}{ccccc}
& 650 MHz & 1.3 GHz & 2.6 GHz & 3.9 GHz \\
\hline
EP/BCP & N/A & $856 \pm 85 $~nm & $2469 \pm 88 $~nm\footnote{The $2.6$~GHz cavity was the only EP'd, both $1.3$~GHz and $3.9$~GHz cavities were instead BCP'd.} & $856 \pm 85 $~nm\footnote{No ``$f$ vs $T$'' data was acquired, the mean-free-path value is assumed equal to that measured at $1.3$~GHz.}\\
$\SI{120}{\degreeCelsius}$ baking\footnote{Value obtained from low-energy $\mu$SR\cite{Morenzoni_PRL_1994} measurements on cavity cut-outs\cite{Romanenko_AppPhysLett_2014}.} & $16 \pm 8 $~nm & $16 \pm 8 $~nm & $16 \pm 8 $~nm & $16 \pm 8 $~nm\\
2$/$6 N-doping & $80 \pm 10 $~nm & $122 \pm 3 $~nm & $96 \pm 4 $~nm & $116 \pm 3 $~nm\\
\end{tabular}
\end{ruledtabular}
\end{table*}

The frequency dependence of the sensitivity can be understood by studying the complex vortex resistivity\cite{Checchin_SUST_2017}:
\begin{equation}
\begin{split}
\rho(l,\omega) &\simeq\dfrac{\omega\phi_0^2}{\pi\xi_0^2\left[\left(p(l)-M(l)\omega^2\right)^2+\left(\eta(l)\omega\right)^2\right]}\\
&\quad\times\left[\eta(l)\omega+i\left( p(l)-M(l)\omega^2\right)\right]\text{,}
\end{split}
\label{eq:resistivity}
\end{equation}
where $\phi_0$ is the flux quantum, $\xi_0$ is the coherence length, $M$ and $\eta$ are the vortex inertial mass and drag coefficient\cite{Bardeen_PR_1965}, and $p$ is the pinning constant as defined in Ref.~[\onlinecite{Checchin_SUST_2017}].

Taking the real part of the resistivity ($\rho_1$) and neglecting the vortex inertial mass since $M\approx0$, we can define two limits: i) small frequencies ($\omega\ll p/\eta$) for which $\rho_1\approx\eta\omega^2/p^2$ and ii) large frequencies ($\omega\gg p/\eta$) where $\rho_1$ is constant. These two limits correspond to the regimes also encountered for small ($p\gg\eta$) and large ($p\ll\eta$) mean-free-path values: pinning and flux-flow, respectively as discussed in our previous work\cite{Checchin_SUST_2017}.

\begin{figure}[b]
\centering
\includegraphics[scale=1]{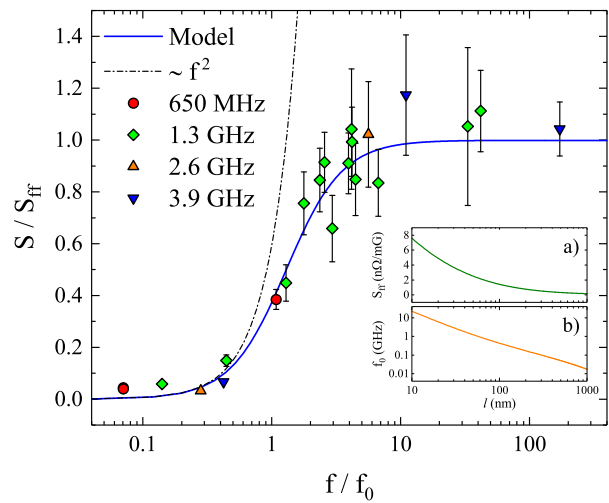}
\caption{\label{fig:All}Normalized sensitivity as a function of normalized frequency data for all the cavities studied in this and in our previous\cite{Martinello_APL_2016} studies. In the inset a), sensitivity in the flux-flow regime as a function of the mean-free-path. In the inset b), depinning frequency as a function of $l$.}
\end{figure}

In order to compare the experimental data to the theoretical model, we normalize the sensitivity with respect to the flux-flow value ($S_{ff}$) and frequency with respect to the depinning frequency\cite{Gittleman_PRL_1966} ($f_0=p/\eta$), both calculated for the mean-free-path value extrapolated for that data point. In Fig.~\ref{fig:All}, the normalized sensitivity ($S/S_{ff}$) is plotted against the normalized frequency ($f/f_0$) for all data acquired in this and in the previous\cite{Martinello_APL_2016} studies.

In the inset a) of Fig.~\ref{fig:All}, we report the flux-flow sensitivity as a function of the mean-free-path calculated using the definition of resistivity in the flux-flow regime:
\begin{equation}
\rho_1(l)\simeq\dfrac{\phi_0^2}{\pi\xi_0^2\eta(l)}\text{.}
\label{eq:fluxflow}
\end{equation}

The depinning frequency as a function of the mean-free-path is instead reported in the inset b) of Fig.~\ref{fig:All}. The values of $f_0$ were calculated numerically as the frequency for which the sensitivity equals half of its flux-flow value for the fixed mean-free-path value. 

The EP'd $2.6$~GHz cavity has mean-free-path of approximately $2500$~nm. This implies that the penetration depth ($\lambda$) of niobium is much shorter than $l$ and niobium is behaving in the extreme clean limit\cite{Pippard_ProcRSocLond_1953}, where the formulation of vortex drag coefficient\cite{Bardeen_PR_1965} is too far from validity. Because of that the EP'd $2.6$~GHz data point will not be compared with the model and will not be plotted with the other data points in Fig.~\ref{fig:All}.

As shown in Fig.~\ref{fig:All}, the experimental data extrapolated at zero accelerating field is well described by the theoretical model and the trend in agreement with vortex resistance measured for PbIn and NbTa alloys\cite{Gittleman_PRL_1966}. For large frequencies ($f/f_0\gg1$), the sensitivity behaves as in the pure flux-flow regime, and it is not dependent on the frequency anymore. For small frequencies ($f/f_0\ll1$), the sensitivity decreases as the frequency decreases with a quadratic law, as discussed previously. 
\begin{figure}[t]
\centering
\includegraphics[width=8cm]{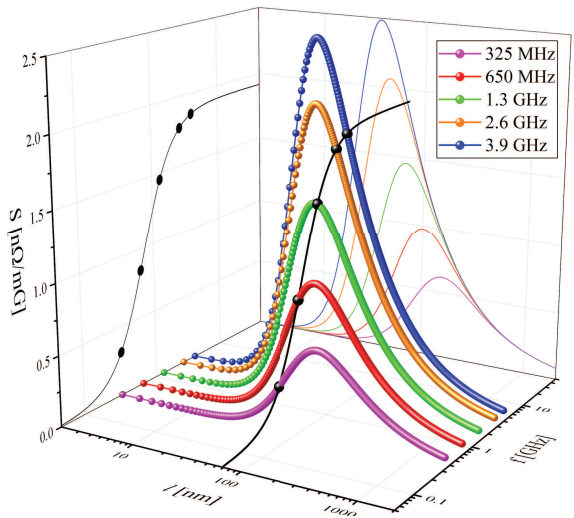}
\caption{\label{fig:3D}Three-dimensional representation of the sensitivity phase space for a fixed pinning point with position $q_0=20$~nm and pinning strength $U_0=1.1$~MeV/m (parameters definition in Ref.~[\onlinecite{Checchin_SUST_2017}]).}
\end{figure}

For the intermediate frequency values falling in between the pinning and flux-flow regimes, the frequency dependence of the trapped flux surface resistance is rather complex and dependent on the mean-free-path and on the number and position of pinning centers interacting with the flux line. Assuming a single pinning point at a fixed distance from the rf surface, we can simulate the sensitivity to trapped flux as a function of both the frequency and mean-free-path, as shown in Fig.~\ref{fig:3D}. The colored points represent $S$ vs $l$ at a fixed frequency, while the black solid line represents $S$ vs $f$ for a fixed mean-free-path ($l=70$~nm). From the solid black line projection on the ``$S\,f$'' plane, it becomes again clear that the sensitivity has a sigmoidal-like trend with $log(f)$, and becomes constant for large $f$. It is also interesting to observe from the projection in the ``$S\,l$'' plane that the peak of $S$ as a function of $l$ moves towards lower mean-free-path values increasing the frequency such that the vortex response at $l=70$~nm transitions from the pinning regime (left side of the peak) to the flux-flow regime (right side of the peak) somewhere in between $650$~MHz and $1.3$~GHz.
\begin{figure}[b]
\centering
\includegraphics[scale=1]{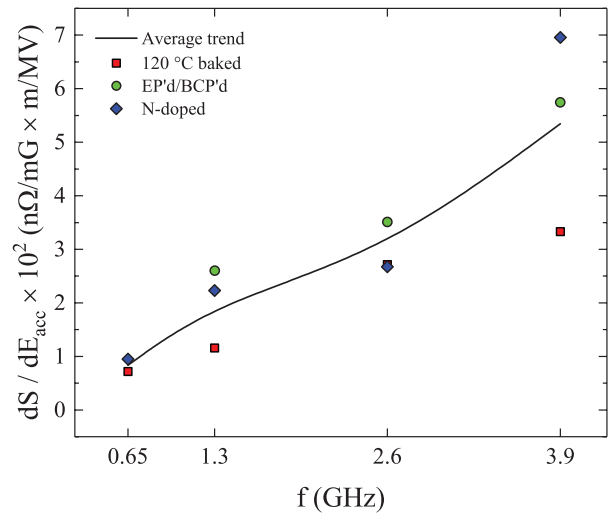}
\caption{\label{fig:Slope}Angular coefficient of the sensitivity to trapped flux as a function of the accelerating gradient, from the experimental data reported in Fig.~\ref{fig:Data}.}
\end{figure}

Let us consider Fig.~\ref{fig:Data} again. It can be observed that the sensitivity increases as a function of the accelerating gradient, as previously observed in 1.5~GHz niobium on copper and 1.3~GHz bulk niobium cavities\cite{Martinello_APL_2016,Benvenuti_PhysicaC_1999}. Even if the model presented here is valid only for low accelerating fields owing to the linear pinning response approximation adopted, we can take advantage of it in order to qualitatively describe the sensitivity field dependence. By increasing the rf field amplitude, the rf screening currents increase accordingly, leading to larger vortex displacements at the pinning center. We should then expect the trapped flux surface resistance to increase with the accelerating field, since the wider the oscillation, the higher the dissipation, as experimentally observed.

Under the same assumptions, we should expect that the slope of the sensitivity as a function of the accelerating gradient ($dS/dE_{acc}$) should increase with the operational frequency. By simple reasoning, we should expect that the power dissipated by the drag force in the flux-flow regime is $\sim\eta v^2$ (with $v$ the vortex velocity). This implies that by fixing the field amplitude\textemdash maximum vortex displacement from the pinning center\textemdash the vortex must move faster as a function of $\omega$ to complete its oscillation inside one rf period. This translates to a faster increasing of the dissipated power as a function of $E_{acc}$ for higher frequencies, and hence, to a higher $dS/dE_{acc}$.

In Fig.~\ref{fig:Slope}, we report the slope of the sensitivity data as a function of the cavity frequency. The $dS/dE_{acc}$ values were extrapolated by fitting the experimental data reported in Fig.~\ref{fig:Data} with a linear regression. As expected, $dS/dE_{acc}$ increases systematically with the frequency; no clear dependence as a function of the surface treatment nor as a function of the mean-free-path is observed. 

In conclusion, we studied the frequency dependence of the trapped flux surface resistance of bulk niobium SRF cavities as a function of the accelerating gradient. We demonstrated that $S$ increased as a function of the frequency with a complex dependence on $\omega$. In the low accelerating gradient limit, $S$ has a $~\omega^2$ dependence in the pinning regime ($f/f_0\ll1$) and is frequency-independent in the flux-flow regime ($f/f_0\gg1$), as expected from our theoretical description\cite{Checchin_SUST_2017}. We also showed that the sensitivity has an accelerating field dependence function with the rf frequency: the higher the frequency, the steeper the field dependence of $S$.

This study is of practical interest for SRF and accelerator physics research communities as it provides guidelines for future cryomodules design; it is also important for the phenomenological understanding of the behavior of SRF cavities because it offers important insight on the physics of vortex dissipation under rf fields. 

This work was supported by the United States Department of Energy, Offices of High Energy and Nuclear Physics and by the DOE HEP Early Career grant of A.~Grassellino. Fermilab is operated by Fermi Research Alliance, LLC under Contract No. DE-AC02-07CH11359 with the United States Department of Energy.

%

\end{document}